# The impact of generative artificial intelligence on academic development of Chinese students in humanities and social sciences

Lei Fan[1,*] and Fangxue Liu[1, 2]

[1] Design School, Xi'an Jiaotong-Liverpool University, Suzhou, 215123, China

[2] School of Computer Science and Informatics, University of Liverpool, Liverpool, L69 3BX, UK;

* Correspondence: lei.fan@xjtlu.edu.cn

**Abstract:** Generative artificial intelligence (GenAI) is reshaping learning in higher education, with particularly pronounced implications for the humanities and social sciences (HSS), where learning outcomes are commonly expressed through written and interpretive forms that align closely with GenAI's capabilities. Yet, systematic evidence on the educational impacts of GenAI on HSS students remains limited. Addressing this gap, this study draws on a large-scale survey of HSS students in China to examine its role in academic development. Guided by relevant learning theories, this study focuses on four dimensions: patterns of use, effects on learning processes and academic performance, challenges associated with GenAI use, and preferred approaches to curricular integration. We found that more than half perceived enhanced learning motivation, independent thinking and creativity, although a substantial minority reported little change or even decline. Comparatively, a notably larger majority reported academic performance gains, although these gains may partly reflect limitations in conventional assessment practices. The study identifies variations in perceived learning and performance improvements among students with differing durations of GenAI experience, along with observable disciplinary differences and modest gender differences. While an overwhelming majority valued the importance of ethical considerations, only slightly more than half were satisfied with privacy protection. Limited accuracy and overreliance emerged as the most pressing concerns reported by students. Students favored partial or optional curricular integration supported by practice-oriented training, and widely recognized GenAI's significance for their future professional development. Grounded in student perspectives, this study offers evidence-based recommendations for the responsible and pedagogically meaningful integration of GenAI.

## 1. Introduction

The rapid advancement and widespread adoption of generative artificial intelligence (GenAI) are reshaping knowledge production and learning practices in higher education (Qian, 2025). With the emergence of accessible GenAI-powered tools, university students now integrate GenAI into a wide range of academic tasks, including summarizing readings, drafting assignments, clarifying complex concepts, and organizing research ideas (Almassaad et al., 2024). Beyond these practical applications, research highlights several key benefits, such as improved learning efficiency, immediate feedback, personalized support, and reduced barriers to academic participation(Chan & Hu, 2023; Qian, 2025). These affordances position GenAI as a transformative educational technology that is influencing learning strategies, assessment practices, and students' perceptions of academic competence (Chan & Hu, 2023; M. Kim & Adlof, 2024).

Humanities and social sciences (HSS) prioritize interpretive understanding, critical argumentation, originality, and the development of an individual intellectual voice (Karjus, 2025). Accordingly, learning outcomes are often evaluated through the quality of reasoning, interpretive coherence, and creative expression, rather than through standardized or technically verifiable answers. From the constructivist perspective (Piaget, 1964, 2005; Vygotsky & Cole, 1978; Zuengler & Miller, 2006), HSS learning is understood as a dialogic process of meaning-making grounded in reflection and interpretation. Within this framework, GenAI functions as a mediating tool that can either scaffold the learning processes or bypass them. The latter may undermine core HSS disciplinary values. Consequently, examining how HSS students employ GenAI in learning represents a central pedagogical challenge.

Although a growing body of research has examined students' behaviors and attitudes toward GenAI in higher education (Almassaad et al., 2024; Chan & Hu, 2023; Jaboob et al., 2025), large-scale and systematic empirical studies focusing specifically on HSS students remain limited. In addition, the sociocultural context of higher education in China adds a further layer of complexity. Chinese universities have undergone rapid transformation, accompanied by strong institutional emphasis on efficiency, performance, and technological innovation (Gao & Wang, 2022). At the same time, academic integrity, originality, and ethical technology use remain central concerns for educators and policymakers. How Chinese HSS students navigate these potentially competing expectations in their engagement with GenAI remains underexplored.

To address these gaps, this study investigates the educational impacts of GenAI on Chinese HSS students using a large-scale questionnaire survey, aiming to provide empirical evidence to inform educators, institutions, and policymakers on the responsible and effective integration of GenAI into HSS education. It systematically examines how GenAI shapes HSS students' learning practices and perceptions in China. Specifically, the study addresses the following research questions:

- In what ways do Chinese HSS students use GenAI in their academic learning?
- How do HSS students in China perceive its impact on learning processes, cognitive engagement, and academic performance?
- What challenges and expectations do Chinese HSS students have regarding the integration of GenAI in HSS education?

The remainder of the article is organized as follows: Section 2 reviews current pedagogical research related to GenAI. Section 3 details the survey design and sampling. Section 4 presents the reliability and validity analysis of the collected data. Section 5 reports and analyzes the survey results. Section 6 outlines the theoretical and practical implications, study limitations, and future research directions. Section 7 concludes the study.

## 2. Literature review

Built on architectures such as Transformers, GenAI systems can produce text, images, and videos by modeling large scale data and increasingly approximating human-like reasoning and expression (Fui-Hoon Nah et al., 2023; Lv, 2023). These capabilities have positioned GenAI as a transformative technology across academic and professional domains (Fui-Hoon Nah et al., 2023). Its applications include personalized learning in education (Chan & Hu, 2023), intelligent content creation in marketing (Kshetri et al., 2024), automated reporting in finance (Rahul Modak, 2025), virtual assistance in healthcare (Samala & Rawas, 2024), and advanced prototyping in art and design (Fathoni, 2023).

In education, GenAI supports personalized and adaptive learning by facilitating idea generation, scaffolding complex tasks, and providing timely feedback, particularly in higher education (Qian, 2025). Empirical studies consistently show that students perceive GenAI as enhancing learning efficiency and reducing cognitive effort. For example, GenAI assisted translation tools have been found to accelerate language acquisition (Wang, 2024), while students using ChatGPT reported improved task efficiency (Sen & Deng, 2025). In addition, research indicates that perceived usefulness and enhanced autonomy can significantly influence students' willingness to adopt GenAI for academic tasks (W. Li et al., 2024; Tbaishat et al., 2026). These findings are consistent with the technology acceptance model (TAM) (Davis, 1989) and the self-determination theory (Deci & Ryan, 2013).

Recent research further suggests that the educational value of GenAI extends beyond improving short-term productivity to supporting deeper cognitive development. Studies conducted in an "Introduction to Linguistics" writing course demonstrated that GenAI, when used as a scaffolded writing assistant, can enhance students' academic expression and interpretive abilities (Wang & Ren, 2024). Beyond writing-related skills, evidence indicates that GenAI can positively influence higher-order thinking, particularly problem-solving and critical thinking abilities (Zhao et al., 2025; Li et al., 2024). Research on creativity has likewise shown that GenAI may stimulate creative performance by offering diverse modes of expression and generating novel ideas for exploration (Alzubi et al., 2025). However, emerging evidence also highlights limitations in GenAI's ability to support advanced cognitive outcomes independently, especially in areas such as critical reasoning and argument development (Qian, 2025; Farrokhnia et al., 2026). Researchers therefore emphasize that the effectiveness of GenAI depends heavily on instructional design. For instance, inquiry-oriented and scaffolded pedagogical approaches appear particularly effective in leveraging GenAI to promote metacognitive regulation, argumentative reasoning, and creative generation (Li et al., 2026).

Numerous studies suggest that the educational impact of GenAI is shaped by how it alters student engagement with learning. When learners remain cognitively active, GenAI can function as external scaffolding that supports engagement with complex content and promotes independent problem solving (H. Kim et al., 2025; Qian, 2025). This perspective aligns with the constructivist theory (Piaget, 1964, 2005), which views learning as an active process of knowledge construction, and with the dialogic learning theory (Bakhtin, 1981), in which GenAI serves as a dialogic learning resource. When students interrogate, adapt, or critique GenAI generated outputs, GenAI can support exploration, conceptual understanding, and navigation of disciplinary conventions and debates (Tang et al., 2024).

At the same time, scholars caution that uncritical reliance on readily available GenAI generated content may undermine deep learning and original intellectual engagement (Fui-Hoon Nah et al., 2023). The fluency and apparent authority of such outputs can create an illusion of understanding, reduce involvement in conceptual framing and argumentative reasoning, and limit investment in cognitive processes (Fui-Hoon Nah et al., 2023), which runs counter to the principles of self-regulated learning (Zimmerman & Schunk, 1994). These concerns are particularly pronounced in HSS disciplines, where interpretive judgment, subjective reasoning, and higher-order cognitive skills such as critical thinking, creativity, and argumentation are central learning outcomes (Karjus, 2025; J. Li & Qi, 2025). In this context, GenAI's ability to produce polished academic outputs complicates the distinction between pedagogical support and cognitive substitution.

Consequently, ethical and normative concerns constitute a central strand of the pedagogical literature related to GenAI. Existing studies highlight challenges related to academic integrity, responsible authorship, and attribution when students incorporate GenAI generated content into academic work (Yusuf et al., 2024). These issues demand clear pedagogical and institutional guidance that defines acceptable uses of GenAI while supporting disciplinary learning goals (Haroud & Saqri, 2025).

AI-enabled feedback represents another extensively studied theme. Research indicates that GenAI-based feedback systems can address long standing challenges of scalability and timeliness, particularly in writing assessment and grading (Escalante et al., 2023). While these GenAI systems are effective in providing feedback, their limitations in fostering deeper cognitive engagement, such as critical reasoning and argument development (Farrokhnia et al., 2026; Qian, 2025), have led researchers to recommend hybrid feedback models that combine GenAI-generated content with human expertise (Banihashem et al., 2025; Escalante et al., 2023).

## 3. Methodology

### *3.1. Questionnaire Design*

This study utilized an anonymized questionnaire survey to investigate the use of GenAI among Chinese HSS students. The questionnaire comprised 21 scale-based questions to identify students' frequently used scenarios, perceived impacts, and attitudes, 5 multiple-choice questions to capture specific behaviors or preferences, and multiple demographic questions to collect information on gender, institution location, major, educational levels, and duration of GenAI use. The scale-based questions were organized into four thematic areas: (1) scenarios of GenAI use, (2) perceived impacts of GenAI on learning efficiency, active learning motivation, independent thinking, and creativity, (3) challenges and concerns in GenAI adoption, and (4) expectations and perspectives on future applications of GenAI in HSS disciplines.

### *3.2. Sampling strategy*

The study protocol was approved by the University Research Ethics Review Panel. The survey was conducted in accordance with relevant ethical guidelines and regulations, and informed consent was obtained from all participants.

A combination of purposeful and convenience sampling was employed. Purposeful sampling aimed to ensure coverage across different HSS disciplines, institution levels (double first class and general institutions) and types (universities and vocational colleges), and educational levels (undergraduates and postgraduates), while convenience sampling allowed broad participation from accessible student populations.

Data was collected anonymously in January 2025 using Wenjuanxing, an online survey platform. Approximately 1000 responses were received. Following data screening and the removal of invalid responses (primarily from non-target majors such as engineering and technology), the final valid sample consisted of 915 students from HSS majors. As shown in Table 1, participants represented a broad range of HSS disciplines in China. The disciplinary distribution shows that students from education-related majors formed the largest proportion of the sample, followed by economics and management, arts, and law. Smaller proportions of respondents were drawn from literature, history and sociology, psychology, and other HSS-related fields.

The sample was predominantly female, with 736 female participants (80.44%) and 179 male participants (19.56%). This gender imbalance reflects the specific demographic structure of HSS students in Chinese universities, where much more female students are enrolled in HSS majors due to factors such as societal gender norms, differential encouragement during earlier stages of education, and perceived labor market suitability.

Regarding educational levels, the majority of respondents were undergraduates (86.01%), while postgraduates accounted for 128 participants (13.99%). This distribution mirrors the overall structure of HSS enrollment in China, where undergraduate students substantially outnumber postgraduate students.

## 4. Reliability and validity assessment of the survey data

The internal consistency of the questionnaire was evaluated using Cronbach's α coefficient (Cronbach, 1951), a widely accepted indicator of the extent to which items within a scale measure the same construct. Cronbach's α ranges between 0 and 1, with larger values reflecting stronger inter-item correlations. In our study, a total of 21 Likert-scale items were subjected to reliability testing using SPSS® Statistics Version 29.0. The analysis produced Cronbach's α coefficients exceeding the recommended threshold (i.e., 0.8) for excellent reliability across all dimensions, with an overall α value of 0.912, demonstrating a high level of internal consistency among the survey items.

To further examine the construct validity of the instrument, the Kaiser–Meyer–Olkin (KMO) measure of sampling adequacy (Kaiser, 1974) and Bartlett's test of sphericity (Tobias & Carlson, 1969) were conducted. The KMO statistic evaluates whether the data is suitable for factor analysis, with values above 0.90 regarded as excellent. As shown in Table 2, the KMO value of 0.927 confirms excellent sampling adequacy. Bartlett's test assesses whether the correlation matrix significantly differs from an identity matrix; a significant result ($p < 0.05$) indicates that the variables are sufficiently correlated for factor analysis. Table 2 shows that the Bartlett's test yielded a highly significant result ($p < 0.001$), leading to the rejection of the null hypothesis and verifying that factor analysis is appropriate. Collectively, these findings indicate that the questionnaire demonstrates satisfactory internal consistency and supports the appropriateness of subsequent analyses.

**Table 1.** Basic information on 915 participants, including their specific disciplines, gender and educational levels.

| Category | Subcategory | Count/Percentage |
|---|---|---|
| HSS disciplines | Education | 442 (48.31%) |
| | Economics and Management | 168 (18.36 %) |
| | Arts | 99 (10.82%) |
| | Law | 94 (10.27%) |
| | Literature | 47 (5.14%) |
| | History and sociology | 33 (3.61%) |
| | Psychology | 10 (1.09%) |
| | Others | 22 (2.40%) |
| Gender distribution | Male | 179 (19.56%) |
| | Female | 736 (80.44%) |
| Educational levels | Undergraduate | 787 (86.01%) |
| | Postgraduate | 128 (13.99%) |

**Table 2.** Results of Cronbach's α coefficient for internal consistency, KMO for measuring sampling adequacy, and Bartlett's test of sphericity.

| **Cronbach's $\alpha$ value** | | 0.912 |
|---|---|---|
| **KMO value** | | 0.927 |
| **Bartlett's test of sphericity** | Chi-square $\chi^2$ | 9872.447 |
| | Degree of freedom Df | 210 |
| | $p$ value | < 0.001 |

## 5. Result

### *5.1. Academic purpose of using GenAI*

#### 5.1.1. Application scenarios of adopting GenAI

As shown in Figure 1a, HSS students reported a wide range of academic activities where they have adopted GenAI, particularly in language-intensive and interpretive tasks central to humanities and social sciences learning. The most widely reported application is assistance with thesis or coursework writing (75.41%), indicating that reliance, if not overreliance, on GenAI for assignment completion has become a common practice. More than two thirds of respondents reported using GenAI to understand and explain theories or concepts (68.74%) and to retrieve relevant literature (68.63%), indicating that GenAI has supported students' active engagement with disciplinary knowledge.

Furthermore, a substantial proportion of students reported using GenAI for data processing and analysis (64.48%) and for summarizing and analyzing textual materials (61.31%). These findings suggest that even in disciplines traditionally grounded in qualitative inquiry, students increasingly rely on GenAI to manage the cognitive demands of large and complex information sets. From a constructivist perspective, GenAI functions as an external scaffold that supports learners in navigating theoretical frameworks, key concepts, and dense scholarly literature (Makransky et al., 2025). In addition, more than half of respondents reported using GenAI for research proposal development and brainstorming, suggesting its role as a dialogic and exploratory resource aligned with dialogic learning theory (Bakhtin, 1981).

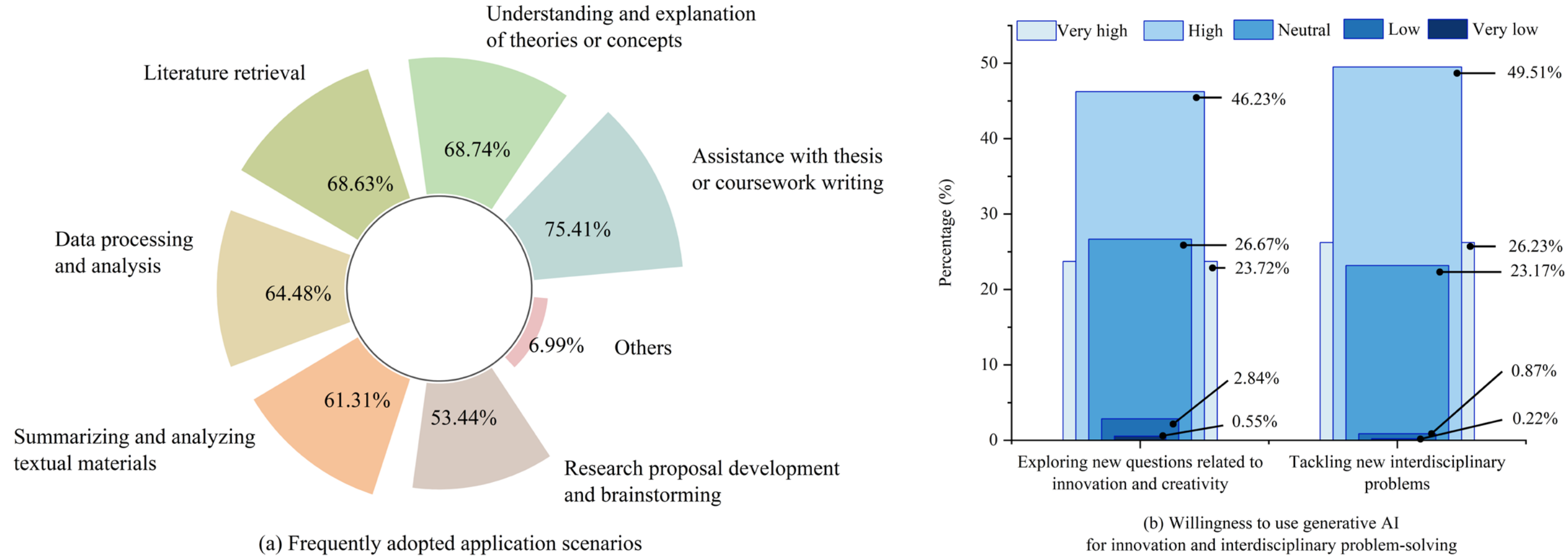

**Figure 1.** Academic purposes of using GenAI tools by HSS students in China: (a) Frequently adopted application scenarios; (b) Willingness to use GenAI for innovation and interdisciplinary problem-solving.

#### 5.1.2. Willingness to use GenAI for innovation and interdisciplinary problem-solving

The survey indicates that students are highly willing to use GenAI in exploratory, innovative, and interdisciplinary contexts, reflecting a strategic mode of engagement. As shown in Figure 1b, nearly 70% of respondents reported high or very high willingness to use GenAI for initial exploration of innovation-related problems. This suggests that HSS students increasingly view GenAI as a cognitive entry point for mapping idea spaces, identifying themes, and generating alternative perspectives in the early stages of creative inquiry. A similar pattern appears in interdisciplinary settings, where GenAI functions as a dialogic resource that supports cross-boundary exploration.

### *5.2. Impact of GenAI on Chinese HSS students' learning and performance*

#### 5.2.1. Perceived impact on learning efficiency

The survey results reveal a near-consensus among HSS students regarding the efficiency-enhancing role of GenAI in their learning activities. As illustrated in Figure 2a, 45.25% of respondents reported a significant improvement in learning efficiency, while a further 50.82% perceived an improvement. In contrast, only a very small proportion reported little change (2.95%) or negative effects (less than 1%).

This widespread perception suggests that GenAI is a highly effective tool for enhancing learning efficiency in HSS contexts. Because these disciplines emphasize extensive literature searching, interpretation, and writing, GenAI's ability to rapidly generate information, provide explanations, and deliver immediate feedback can substantially reduce the time required to complete academic tasks. These efficiency gains are consistent with prior research showing that timely feedback enables students to identify gaps in their understanding and accelerate knowledge construction, thereby improving overall learning efficiency (Qian, 2025).

#### 5.2.2. Perceived impact on active learning motivation

Figure 2a shows that GenAI generally exerts a positive influence on HSS students' motivation for active learning. A total of 65.03% of respondents reported increased motivation, including 23.06% who experienced a significant enhancement. These students may perceive GenAI as an accessible, non-judgmental support that enables self-directed exploration and reduces learning-related anxiety.

Nevertheless, 22.40% reported almost no change, and 12.57% perceived a decline in motivation. This pattern reflects the tension highlighted in self-determination theory (Deci & Ryan, 2013). While GenAI can strengthen autonomy and perceived competence, excessive dependence may weaken intrinsic motivation by shifting attention from learning processes to outcome-oriented task completion. This suggests the importance of guiding students toward reflective and purposeful GenAI use rather than habitual reliance.

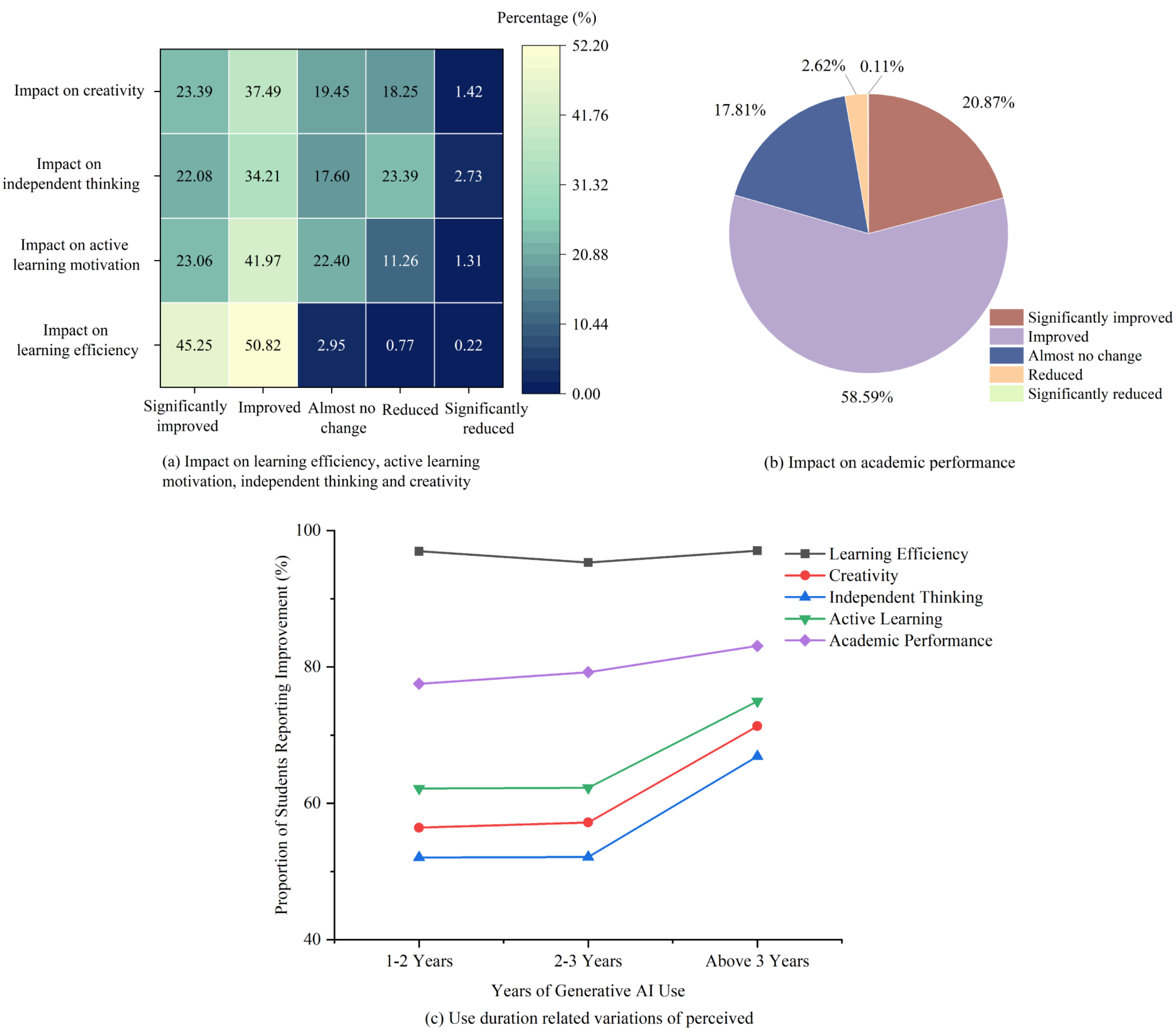

**Figure 2.** Perceived impact of GenAI on Chinese HSS students' learning: (a) Impact on learning efficiency, active learning motivation, independent thinking and creativity; (b) Impact on academic performance; (c) Use duration related variations in perceived improvements across different learning dimensions.

#### 5.2.3. Perceived impact on independent thinking

HSS students' perceptions of GenAI's influence on independent thinking are notably mixed. According to Figure 2a, 22.08% of respondents perceived a significant improvement and 34.21% an improvement, indicating that over half of the students viewed GenAI as supportive of independent thinking. These students may benefit from GenAI-generated alternative perspectives, conceptual clarification, and feedbacks that stimulate critical reflection.

However, a considerable proportion of respondents reported negative effects: 23.39% perceived a decline and 2.73% a significant decline in independent thinking, while 17.60% observed no noticeable change. This polarization highlights the dual role of GenAI in HSS learning. From an epistemic cognition perspective, GenAI can function either as a dialogic partner that encourages critical engagement or as an authoritative source that discourages questioning (Bakhtin, 1981; Tang et al., 2024). When students rely uncritically on fluent and seemingly authoritative GenAI-generated interpretations, opportunities for independent sense-making and reasoning may be reduced.

#### 5.2.4. Perceived impact on creativity

Perceptions of GenAI's impact on creativity among HSS students follow a similarly differentiated pattern as that for independent thinking. As indicated in Figure 2a, 60.88% of respondents reported positive effects on creativity, with 23.39% experiencing a significant improvement. These students may view GenAI as a source of inspiration, idea expansion, or alternative framing of arguments.

At the same time, nearly one-fifth of respondents (19.67%) reported a decline in creativity, and 19.45% perceived little change. This suggests that while GenAI can support creative exploration, it may also constrain originality for some learners. From the standpoint of academic literacies theory, creativity in HSS disciplines is closely tied to voice, interpretation, and epistemic positioning (Alghamdi, 2025; Lea & Street, 1998). In this context, creativity should not be viewed merely as the production of polished work, but as students' capacity to develop original interpretations, construct independent arguments, and express a distinctive intellectual voice. When GenAI-generated text becomes overly influential in shaping expression, students may feel that their personal intellectual contribution is diminished, thereby limiting perceived creativity. Furthermore, because assessment practices in many HSS courses often prioritize coherence, stylistic sophistication, and task efficiency, students may interpret AI-assisted outputs as "creative" even when the underlying processes of idea generation and critical interpretation remain limited.

5.2.5. Perceived impact on academic performance

HSS students' perceptions of academic performance outcomes are predominantly positive. As shown in Figure 2b, 20.87% of respondents reported a significant improvement and 58.58% an improvement, indicating that nearly four out of five students believed GenAI had enhanced their academic performance. Only a small fraction reported no change (17.81%) or negative effects (less than 3%).

However, such performance gains should be interpreted with caution. In this study, academic performance primarily refers to students' perceived success within existing institutional assessment frameworks, such as grades, assignment completion, writing quality, and overall academic efficiency. It does not necessarily indicate deeper conceptual understanding, sustained critical engagement, or long-term intellectual development. Assessment practices in HSS disciplines often emphasize written fluency, coherence, and analytical presentation, areas where GenAI can offer substantial assistance. As a result, perceived improvements may reflect enhanced surface-level outputs rather than meaningful improvements in students' underlying understanding or higher-order cognitive skills. This concern is reinforced by the reported declines in independent thinking and creativity (see Sections 5.2.3–5.2.4). These may suggest a potential disconnect between performance outcomes and actual learning processes for some students.

This disconnection highlights a possible limitation of current assessment strategies, which may be insufficiently equipped to differentiate GenAI-supported task completion from genuine learning in the GenAI era. To address this challenge, HSS assessment practices may need to be recalibrated to place greater emphasis on process-oriented evaluation and cognitive performance assessment. Such approaches may help distinguish students' independent intellectual contributions from improvements that are primarily attributable to GenAI-assisted production.

5.2.6. Effects of GenAI use duration on learning enhancement

Figure 2c presents students' perceived gains across five learning dimensions by their duration of GenAI experience. Learning efficiency shows uniformly high improvement across all experience levels, indicating an immediately strong and stable effect. Consistent with cognitive load theory (Sweller et al., 1998), GenAI quickly reduces the time and effort required for routine academic tasks. These benefits are reported across all duration groups and do not appear to vary substantially with longer use, suggesting they may emerge early and are largely independent of extended experience or explicit pedagogical support. Academic performance likewise demonstrates consistently high perceived improvement (see Section 5.2.5) across duration groups, but follows a more gradual trajectory, with steady gains with longer use.

By contrast, perceived improvements in creativity, independent thinking, or active learning motivation are similar for students with 1–2 and 2–3 years of experience, followed by a sharp increase among those with more than three years. While this pattern may indicate a developmental interpretation, it should be understood that it is based on comparisons between subgroups with different GenAI use durations rather than measured evidence of longitudinal change over time. During the phase of early adaptation, deeper cognitive benefits may remain constrained as learners gradually acquire skills in prompt formulation, critical evaluation of outputs, and integration of GenAI into their own reasoning. From a sociocultural perspective (Vygotsky & Cole, 1978; Zuengler & Miller, 2006), this stage reflects external mediation (Jiang et al., 2025) in which GenAI has not yet been internalized as part of the learner's cognitive repertoire. This process is likely slowed by the limited provision of systematic GenAI-related training by higher-education institutions in China.

Importantly, this delayed growth does not imply weak early effects, because more than half of less-experienced students already reported gains across all dimensions. Rather, it indicates that more extensive experience with GenAI may be associated with stronger and more consistent perceived benefits as students develop more sophisticated GenAI use practices. The heightened gains among students with over three years of experience suggest a possible developmental or cumulative nature of GenAI's educational impact, reinforcing the need for structured institutional support.

5.2.7. Cross-discipline comparisons

Figure 3 compares Chinese HSS students' perceived impacts of GenAI across selected disciplines. Owing to sample size limitations, the analysis is restricted to four fields: education, economics and management, arts, and law. Learning efficiency is excluded from the cross-disciplinary comparison because it receives uniformly high ratings across disciplines (see Figure 2c) and exhibits minimal disciplinary variation.

Clear disciplinary differences emerge across the remaining dimensions. Students in arts and in economics and management consistently reported stronger perceived gains in independent thinking, creativity, and active learning motivation than their counterparts in education and law. For instance, over two-thirds of arts students reported improvements in independent thinking (67.67%) and creativity (72.72%), compared with just over half of students in education and law. Negative perceptions are also more pronounced in education and law, particularly for independent thinking, where approximately one-third of education students and nearly one-quarter of law students reported declines. Although perceptions of academic performance are more convergent, with generally high reported gains across all disciplines, the same disciplinary pattern remains evident.

These differences likely reflect variation in disciplinary epistemologies and assessment regimes. Arts, and economics and management prioritize idea generation, interpretation, and flexible problem-solving, which closely align with GenAI's strengths in brainstorming and exploratory tasks. By contrast, education and law are more normatively structured and place greater emphasis on professional standards, heightening sensitivity to perceived risks to autonomy and critical thinking. The findings suggest the importance of discipline-responsive strategies for GenAI integration in higher education, rather than assuming uniform effects across fields.

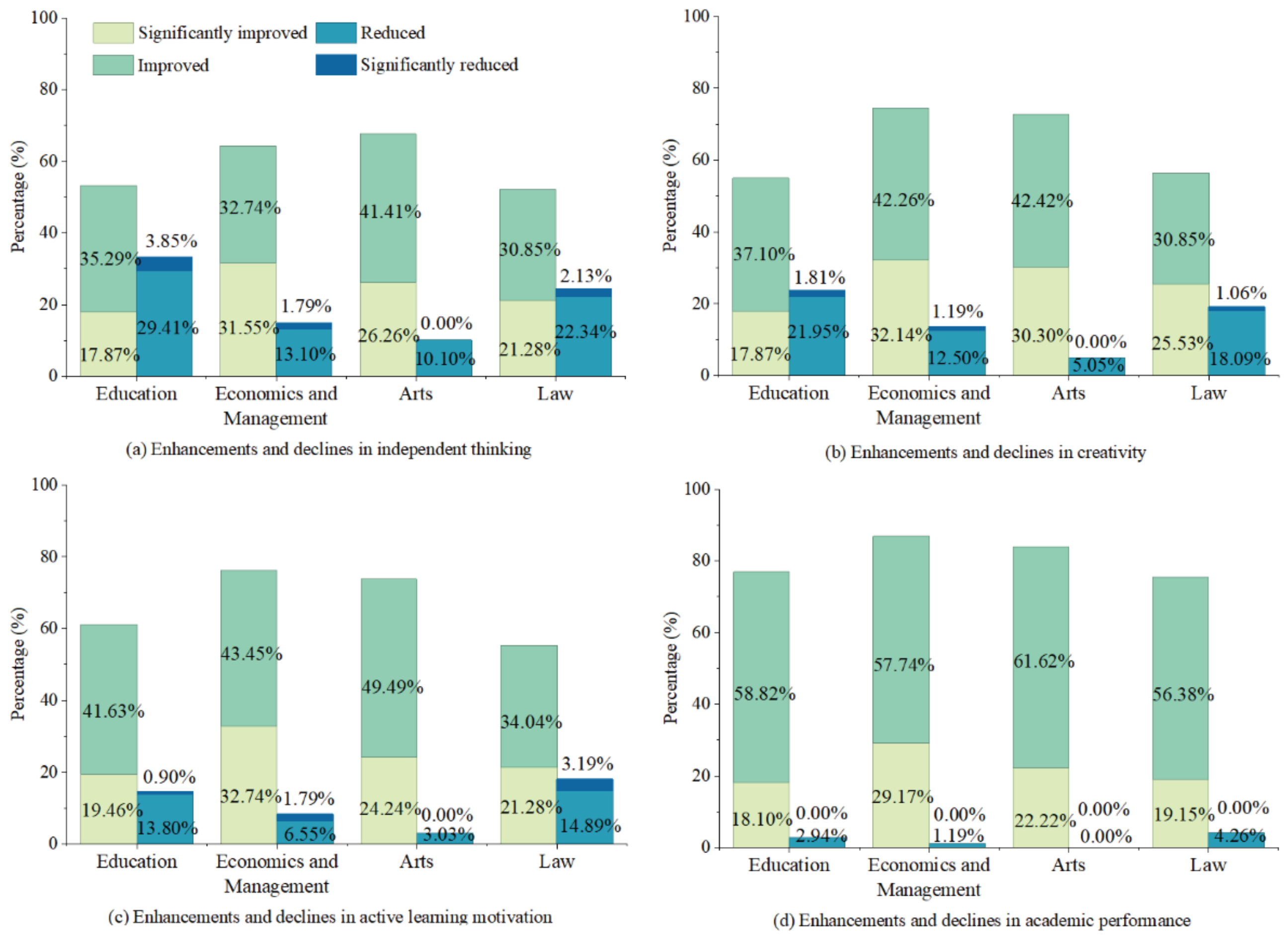

**Figure 3.** Cross-discipline comparison of perceived enhancements and declines in: (a) independent thinking; (b) creativity; (c) active learning motivation; (d) academic performance.

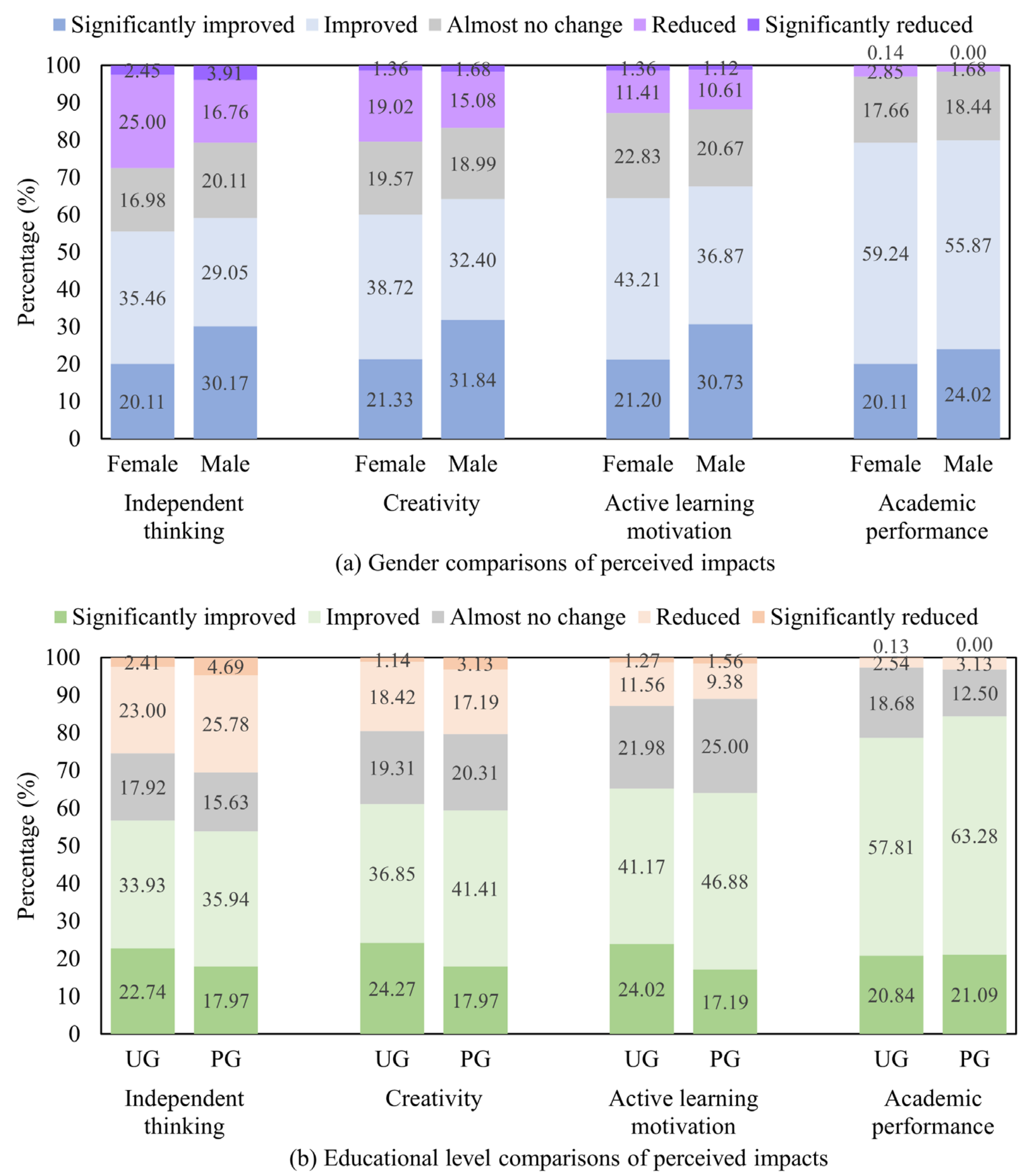

**Figure 4.** Comparison of perceived impacts of GenAI by (a) Gender and (b) Educational level (UG: undergraduate, PG: postgraduate).

5.2.8. Gender differences

Figure 4a shows that male students in HSS disciplines reported higher perceived gains from GenAI in independent thinking, creativity and learning motivation. This aligns with prior research indicating that male students tend to express greater confidence and more positive attitudes toward GenAI adoption, often linked to higher technological self-efficacy and a more instrumental use of digital tools, which enhances perceived learning benefits (Cachero et al., 2025; Matobobo, 2026; Ofosu-Ampong, 2023). By contrast, female students may adopt more evaluative and integrity-focused approaches (Matobobo, 2026), using GenAI in a moderated, supplementary way that tempers perceptions of transformative impact.

Notably, there was very small gender difference in perceived academic performance, in contrast to the gaps observed in cognitive and motivational outcomes. This suggests that while performance gains are broadly shared, perceived cognitive benefits are more sensitive to differences in confidence framing and baseline expectations. Overall, the disparity appears to reflect variations in adoption strategies and evaluative framing rather than differences in learning capacity.

5.2.9. Educational level comparisons

The survey results shown in Figure 4b indicate minimal overall differences between undergraduates and postgraduates in perceived improvements ("improved" and "significantly improved") across cognitive dimensions and learning motivation. Nevertheless, a slightly higher proportion of undergraduates reported "significantly improved" outcomes. This pattern may reflect developmental factors: as students in earlier stages of academic formation, undergraduates may experience GenAI's novelty and dialogic features as more transformative than postgraduates, strengthening their sense of autonomy and exploratory engagement.

Conversely, postgraduates reported slightly greater gains in academic performance. The academic maturity and deeper disciplinary understanding of postgraduates may enable them to more critically and strategically integrate GenAI into complex tasks where GenAI can facilitate more visible and measurable gains. By comparison, undergraduate-level assessments often focus more on foundational knowledge, where academic performance gains from GenAI may appear less direct.

*5.3. Challenges faced by Chinese HSS students*

5.3.1. Key challenges in using GenAI

Although GenAI is widely used and generally viewed as beneficial by HSS students, the survey reveals several serious challenges that complicate the effective and responsible use. The most prominent issue is the limited accuracy of GenAI-generated outputs, identified by 81.42% of respondents (Figure 5), far exceeding all other challenges. This highlights a fundamental tension between students' reliance on GenAI and their trust in its outputs.

The second major challenge is over-reliance on GenAI, reported by 65.36% of respondents. In HSS disciplines, where interpretive agency and argument construction are central (Karjus, 2025), such over-reliance risks shifting learning from meaning-making to output optimization, as well as disengaging themselves from essential learning processes.

Ethical and privacy concerns are also substantial, with 41.20% of students identifying them as major challenges, alongside 36.39% citing insufficient technical support. These concerns point to gaps in institutional policies, guidance, and training amid rapid GenAI adoption. By contrast, challenges related to usability (28.42%) and cost (20.98%) were less prominent, suggesting that most students can access and operate GenAI tools in China.

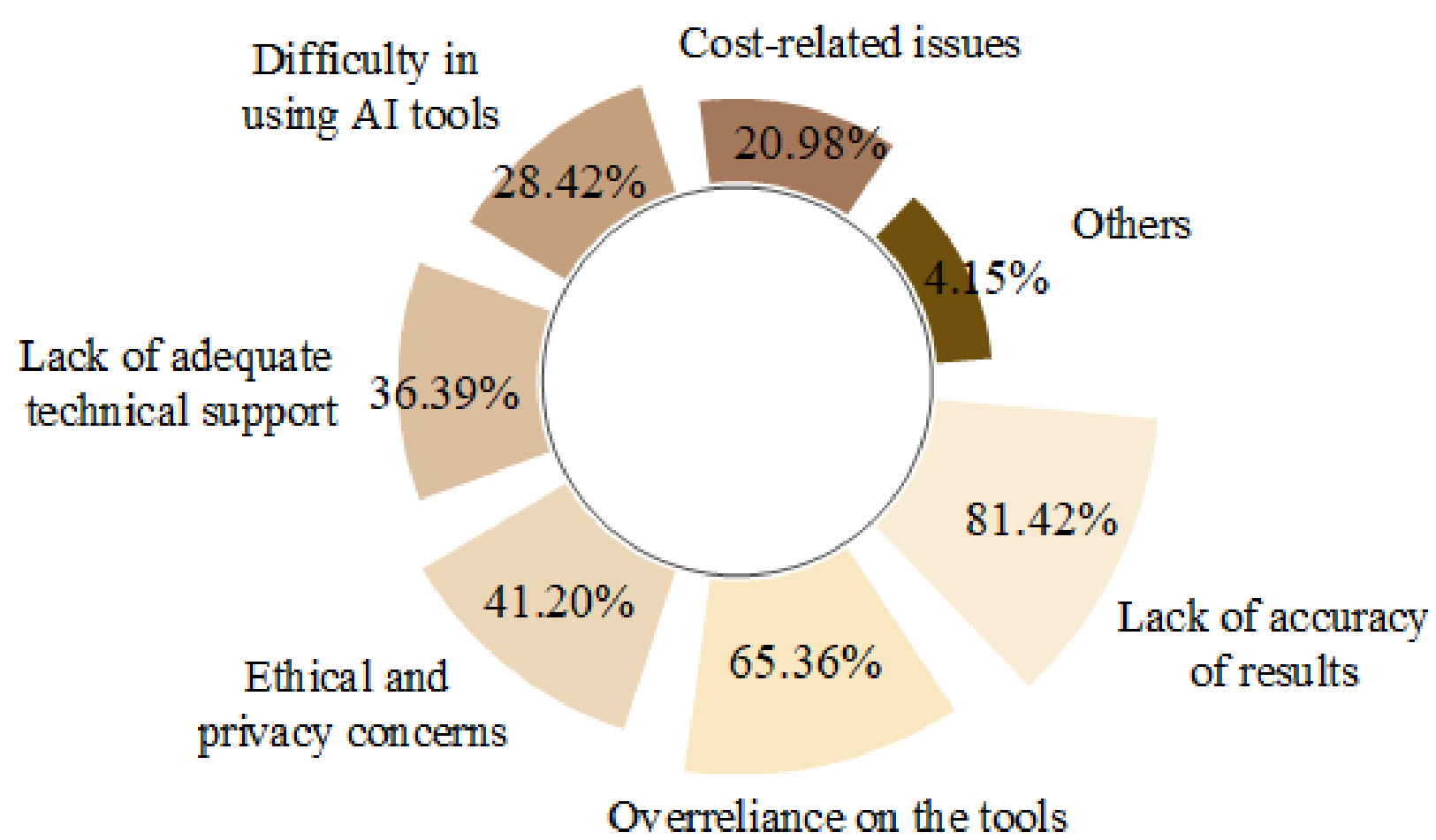

**Figure 5.** Key challenges in using GenAI by HSS students in China.

5.3.2. Adaptability of GenAI in HSS education

As shown in Figure 6a, a majority of students perceived GenAI as aligning with their HSS disciplines, with 65.68% of respondents agreeing or strongly agreeing. This perception likely reflects GenAI's capacity to support core HSS tasks, such as synthesizing literature, explaining concepts, generating ideas, and writing. However, the substantial proportion of neutral responses (31.26%) indicates a degree of ambivalence regarding this disciplinary alignment. This ambivalence may stem from students experiencing GenAI as facilitating surface-level task completion rather than fostering the deeper knowledge construction emphasized in constructivist learning theory (M. Kim & Adlof, 2024).

This uncertainty is further indicated by students' frequent encounters with inaccurate outputs, revealing a structural tension between perceived disciplinary alignment and epistemic reliability. As shown in Figure 6b, 67.10% of students reported encountering inaccurate information frequently or very frequently, while an additional 31.04% reported encountering it occasionally. Although such experiences may partly be influenced by students' varying levels of proficiency with GenAI, they also point to inherent limitations of current GenAI systems. For HSS students whose learning relies heavily on interpretive judgment and engagement with contested, evolving forms of knowledge (Karjus, 2025), this tension is particularly consequential. From an epistemic cognition perspective (Hofer & Pintrich, 1997), repeated exposure to weakly grounded or inaccurate information may erode students' ability to distinguish rigorously justified arguments from superficially plausible claims.

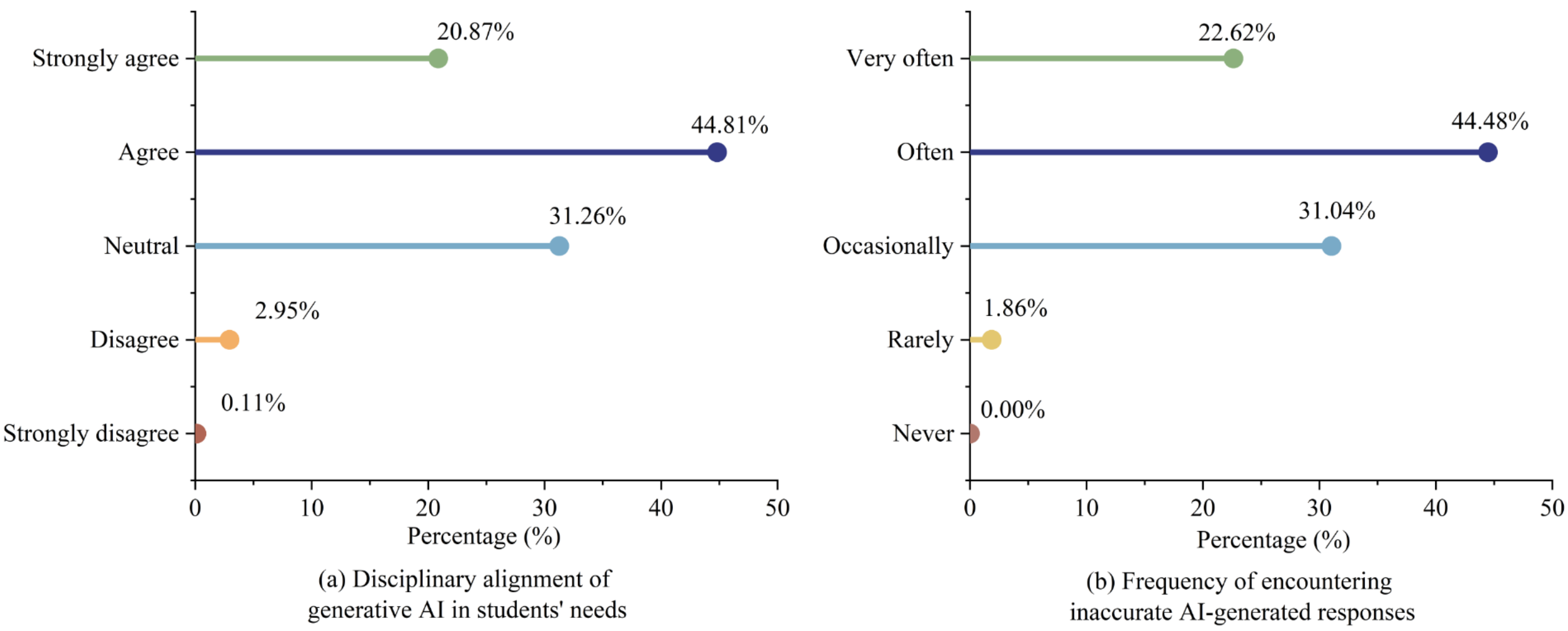

**Figure 6.** Perceived adaptability of GenAI among Chinese HSS students: (a) Disciplinary alignment of GenAI in meeting students' needs; (b) Frequency of encountering inaccurate GenAI-generated responses.

5.3.3. Ethical concerns and data privacy in GenAI use

Ethical considerations emerge as a particularly important matter for HSS students. As shown in Figure 7a, an overwhelming 87.97% of respondents rated ethical issues as important or very important, with nearly half (48.52%) considering them very important. This heightened sensitivity reflects widespread awareness of challenges such as academic integrity, ambiguous authorship, and the blurred boundary between legitimate assistance and inappropriate adoption. These issues are especially pronounced in writing-intensive HSS disciplines. This concern aligns with academic literacies theory (Lea & Street, 1998), which frames reading and writing as a socially situated practice closely tied to identity, knowledge construction, and epistemic authority. When GenAI is capable of producing polished academic prose, students may find it difficult to determine where acceptable scaffolding ends and unethical delegation begins, resulting in ethical uncertainty.

Perceptions of data privacy protection, shown in Figure 7b, are more mixed. While 56.94% of respondents expressed satisfaction or high satisfaction with GenAI tools' privacy performance, a substantial proportion (37.27%) remained neutral, and 5.80% reported dissatisfaction. This cautious stance suggests that even when students continue to use GenAI tools, they may do so with reservations about how their data is stored, processed, or reused. As prior research indicates, unresolved privacy concerns can weaken trust and inhibit sustained engagement, even when perceived usefulness is high (Đerić et al., 2025).

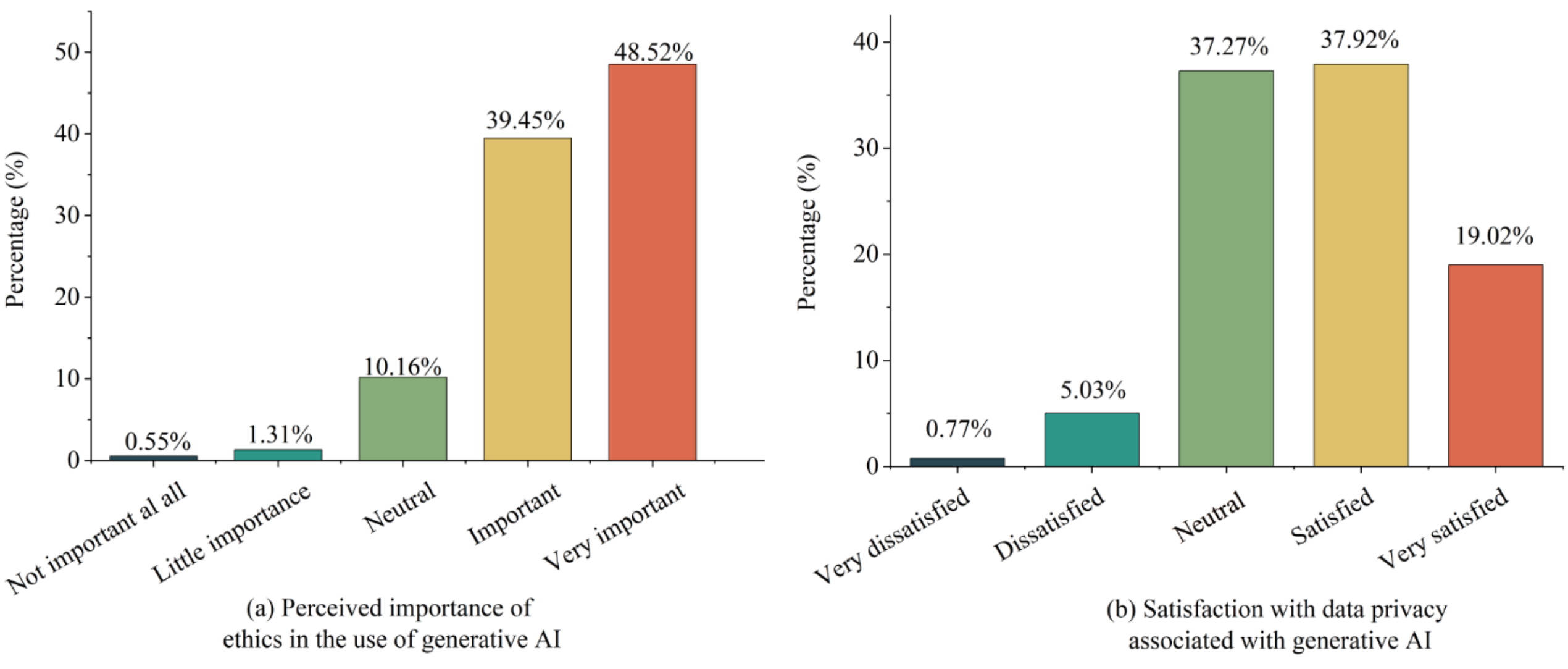

**Figure 7.** Ethical and data privacy considerations among Chinese HSS students: (a) Perceived importance of ethics in the use of GenAI; (b) Satisfaction with data privacy associated with GenAI.

### *5.4. Expectations regarding GenAI use among Chinese HSS students*

#### 5.4.1. Attitudes toward integrating GenAI into HSS education

Survey results indicate cautious support for integrating GenAI into HSS education. As shown in Figure 8a, most respondents preferred partial integration (54.97%), favoring selective use in appropriate courses or learning contexts. This preference indicates a shared understanding that educational value derives from pedagogically purposeful application rather than from the mere presence of technology (M. Kim & Adlof, 2024). An additional 27.21% considered embedding GenAI as a useful but optional tool, while only 17.16% supported full integration to the teaching process.

Students' views on whether GenAI could replace traditional teaching are more divided. As shown in Figure 8b, while 42.30% considered such replacement likely or very likely, 29.51% viewed it as unlikely or impossible, and 28.20% were uncertain. This lack of consensus suggests that, while students acknowledged the growing instructional capacities of GenAI, they did not regard it as a replacement for human-led teaching, which remains essential for dialogic engagement, mentorship, and disciplinary socialization in HSS education.

#### 5.4.2. Expectations for GenAI-related training and institutional support

Students reported strong expectations for institutional support in developing competence in the use of GenAI. As shown in Figure 8c, in-depth practical courses were most strongly preferred (71.04%), followed by basic introductory courses (61.86%) and online formats (47.76%), while lecture-based approaches were viewed less favorably. Overall, these preferences indicate a clear demand for practice-oriented guidance rather than theoretical instruction. From a TAM perspective, such training is likely to enhance perceived usefulness and ease of use, thereby supporting more informed and sustainable adoption of GenAI (Davis, 1989; W. Li et al., 2024).

In terms of governance, Figure 8d indicates broad support for institutional regulation rather than outright prohibition. Most respondents favored the provision of clear usage guidelines (68.09%), strengthened training (67.76%), and tailored policies across courses (58.47%). Nonetheless, a substantial minority (22.51%) supported a complete ban, pointing to ongoing concerns about ethical issues and potential overreliance on GenAI (Babayev, 2025).

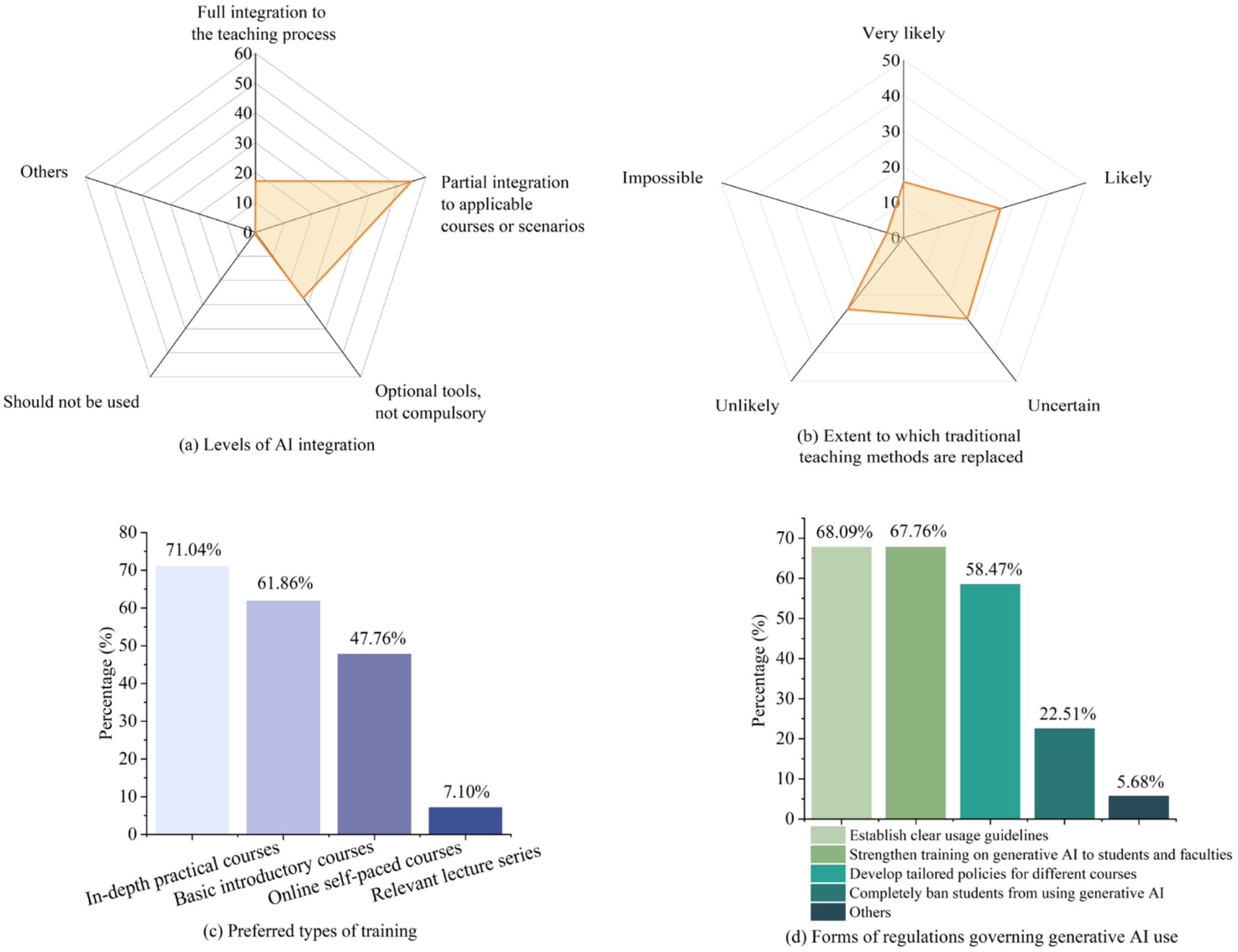

**Figure 8.** Integration of GenAI into HSS education from the perspective of Chinese HSS students: (a) Levels of GenAI integration; (b) Extent to which traditional teaching methods are replaced; (c) Preferred types of training; (d) Forms of regulations governing GenAI use.

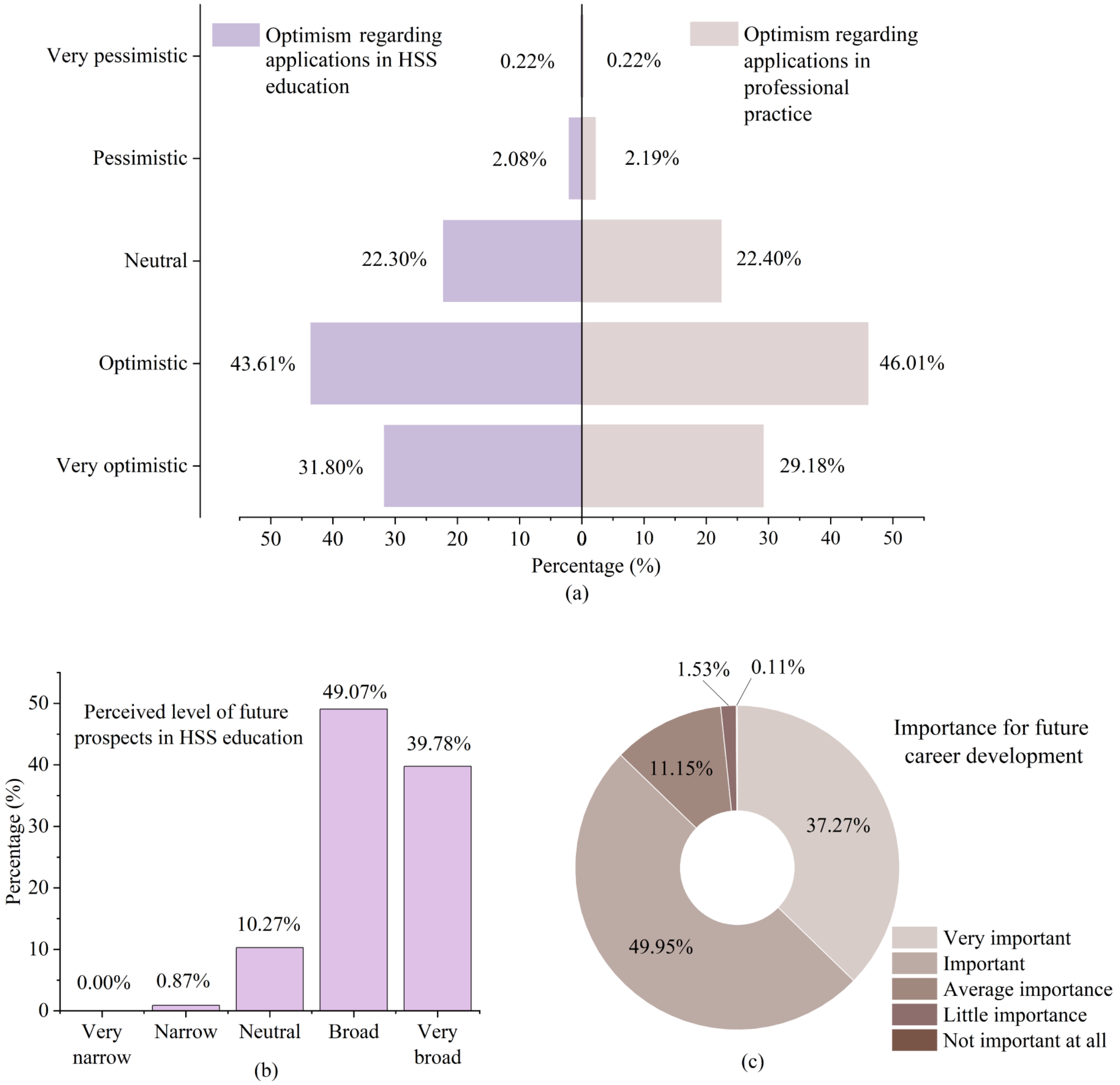

**Figure 9.** Chinese HSS students' perspectives on the prospects of GenAI: (a) Optimism regarding applications in HSS education and professional practice; (b) Perceived level of future prospects in HSS education; (c) Perceived importance for future career development.

5.4.3. Confidence in GenAI's development for HSS education and professional careers

HSS students exhibited strong optimism regarding the future of GenAI in their fields. As illustrated in Figure 9a, 75.41% were optimistic or very optimistic about its development in HSS education over the next five years, with only 2.30% expressing pessimism. Similarly, nearly 89% perceived its potential application in HSS education as broad or very broad (Figure 9b). This optimism likely stems from their direct experiences in using GenAI for HSS-related tasks. Neutral responses (22.30%), however, indicate cautious optimism, likely influenced by ethical concerns, over-reliance, and variable output quality.

GenAI was also seen as highly consequential for future careers. Figure 9c shows that 87.22% rated GenAI as important or very important for career development, suggesting it constitutes career-relevant capital rather than a purely academic tool. Likewise, 75.19% (Figure 9a) expressed optimism about GenAI's role in HSS professional contexts, with pessimism below 3%. These results indicate that GenAI enhances students' sense of instrumental competence by aligning skill development with workplace demands. Yet, the sizable neutral group (22.40%) reflects ongoing uncertainty about GenAI's integration into professional norms and accountability structures.

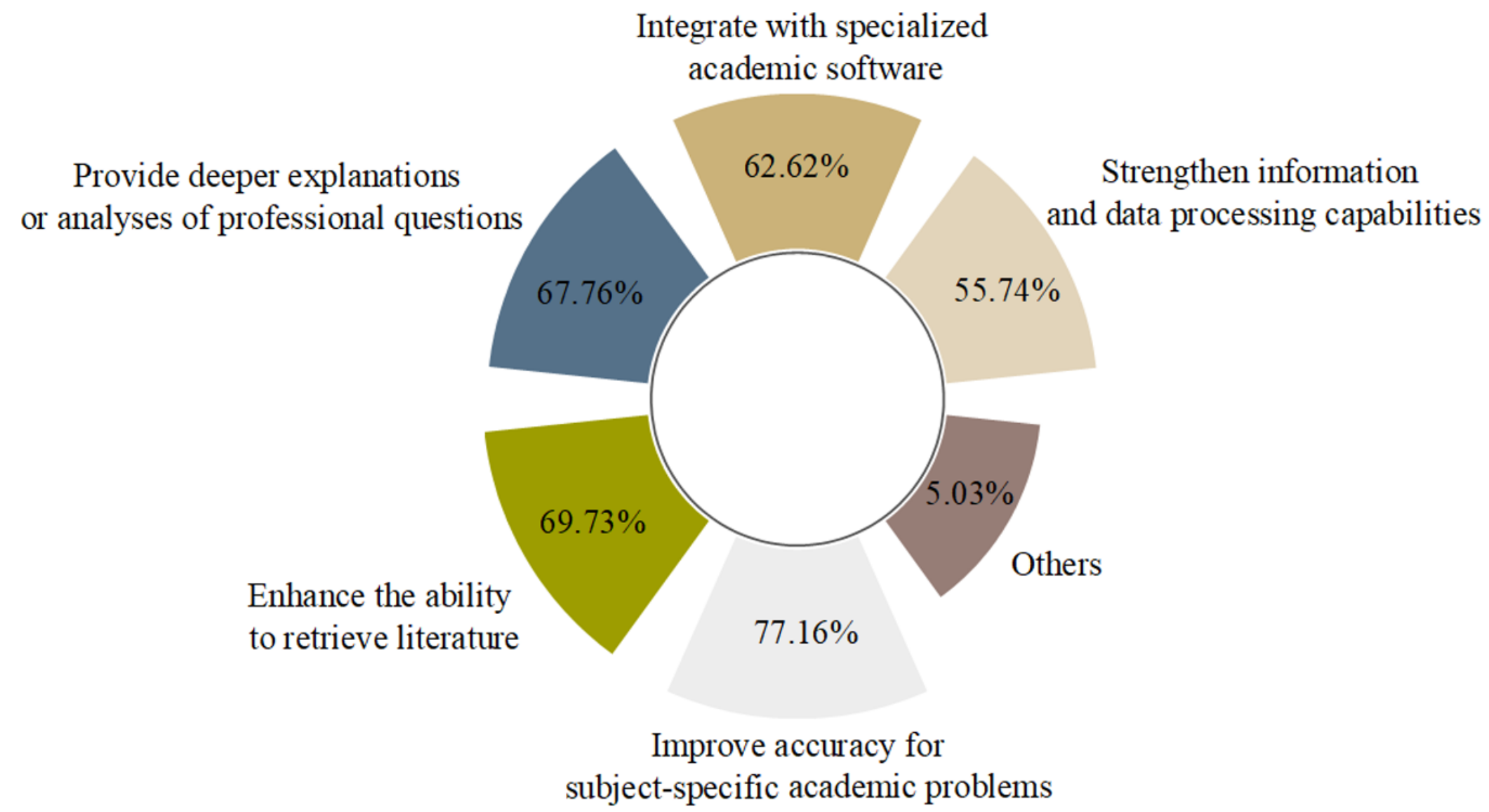

**Figure 10.** Areas of improvement of GenAI in HSS education from Chinese HSS students' perspectives.

5.4.4. Desired improvements in GenAI tools

When asked to prioritize improvements, HSS students most frequently prioritized greater accuracy in discipline-specific tasks (77.16%), enhanced literature retrieval capabilities (69.73%), and more in-depth explanations or analyses of professional questions (67.76%), as illustrated in Figure 10. A substantial portion of students also expressed interest in integration with disciplinary software (62.62%) and stronger data-processing functionalities (55.74%).

These results suggest a clear demand for academically comprehensive and versatile GenAI systems rather than generic conversational tools. From constructivist and epistemic cognition perspectives (Hofer & Pintrich, 1997; Piaget, 1964, 2005), high accuracy and deeper explanatory capacities are critical for fostering genuine understanding, rather than merely facilitating superficial task completion. The emphasis on literature retrieval and analytical support further highlights the central role of scholarly dialogue and evidence-based reasoning in HSS learning contexts.

## 6. Discussion

*6.1. Implications for theory*

The study findings demonstrate that GenAI has almost universally improved the learning efficiency of Chinese HSS students, regardless of their prior GenAI use experience or disciplines (Sections 5.2.1, 5.2.6, and 5.2.7). Reinforcement theory (Skinner, 1981) may clarify this outcome, as the immediate feedback provided by GenAI accelerates iterative learning and encourages sustained engagement. Such evidence also lends empirical support to cognitive load theory (Sweller et al., 1998), as the automation of preparatory tasks appears to reduce extraneous cognitive load and free cognitive resources for deeper processing.

Effects on independent thinking and creativity are more complex (Sections 5.2.3 and 5.2.4). While many students reported gains, roughly one-quarter perceived declines in critical thinking, and nearly one-fifth reported reduced creativity. Gains in learning efficiency may undermine these cognitive skills when GenAI is used irresponsibly. Multiple theoretical perspectives help interpret these mixed effects. Academic literacies theory (Lea & Street, 1998) indicates that over-reliance (a widely recognized challenge by surveyed respondents) on GenAI may limit students' personal voice and interpretive agency, thereby constraining creative expression. Self-regulated learning theory (Zimmerman & Schunk, 1994) further highlights that excessive dependence can undermine deep learning, producing superficial achievement and reducing intrinsic motivation (empirically reflected in the observed gap between students' perceived gains in academic performance and their active learning motivation in this study). In addition, epistemic cognition theory (Hofer & Pintrich, 1997) suggests that GenAI can function either as a dialogic partner that stimulates critical reflection or as an authoritative source that suppresses questioning.

However, the study also sheds light on the longitudinal trajectory of GenAI-assisted development in creativity and independent thinking (Section 5.2.6). While gains in efficiency appear almost immediately, improvements in deeper cognitive skills emerge more gradually, with notable increases reported after approximately three years of GenAI use. This pattern suggests a promising trend toward the responsible integration of GenAI, as students increasingly engage with it as a dialogic partner and develop more sophisticated practices for incorporating it into their learning. From a sociocultural perspective (Vygotsky & Cole, 1978), this progression can be understood as a shift from external mediation to internalization. Initially, GenAI functions primarily as a convenience tool. Over time, however, sustained and reflective engagement allows students to integrate it into their own cognitive processes.

### *6.2. Implications for practice*

GenAI has substantial potential for HSS education, but effective integration requires addressing several interconnected challenges. Accuracy of GenAI outputs is a primary concern. More than 80% of students identified accuracy as critical, and around two-thirds reported frequent exposure to errors (Sections 5.3.1–5.3.2). In HSS disciplines, these errors can distort reasoning, weaken conceptual understanding, and undermine disciplinary discourse. Addressing this challenge requires both technical improvements in GenAI systems and the systematic development of students' critical evaluation and epistemic judgment skills.

Ethical and data privacy issues constitute another critical concern (Section 5.3.3). Given the authorship-sensitive nature of HSS disciplines, these concerns are closely tied to academic integrity and appropriate use of GenAI. The fact that 22.51% of respondents supported a complete ban on GenAI reflects persistent anxiety among a minority about fairness, accountability, and learner autonomy. Institutions should instead adopt balanced frameworks that provide clear guidelines, ethical education, and opportunities for reflective practice.

Assessment practices also require careful attention. Students reported strong perceived improvements in academic performance (Section 5.2.5), but these gains may partly reflect limitations in current assessments, which often cannot distinguish GenAI-assisted task completion from genuine learning. This highlights the institutions' need to redesign assessment strategies that prioritize reasoning, process, and higher-order understanding, as well as explicit criteria for evaluating originality and critical engagement. Complementary measures such as oral defenses or in-class assessments may further enhance the validity of performance assessment.

Students' preferred modes of training inform effective institutional implementation: they favored learning designs that integrate foundational conceptual instruction with hands-on, practice-oriented experiences (Section 5.4.2). Accordingly, scaffolded learning sequences that progress from conceptual understanding to guided application can foster reflective and sustained engagement with GenAI in HSS education. In addition, our study findings highlight the need for discipline-responsive strategies for GenAI integration in higher education, as marked differences across HSS disciplines (Section 5.2.7) caution against assuming uniform impacts or pedagogical approaches. Students also perceived GenAI as a catalyst for career readiness (Section 5.4.3). Educators can capitalize on this potential by designing curricula that frame GenAI as a tool for professional skill development.

### *6.3. Comparative insights between HSS and engineering students*

Comparing HSS students in this study with engineering students reported in a previous study (Fan et al., 2025) reveals both shared patterns and pronounced disciplinary differences in GenAI use and attitudes. Although both groups reported substantial gains in learning efficiency, HSS students showed a higher level of consensus regarding efficiency improvement. This likely reflects the strong alignment between GenAI and the language-intensive, interpretive, and writing-centered tasks that dominate HSS learning and assessment. Perceived gains in academic performance were also more pronounced among HSS students, whereas engineering students reported a larger gap between efficiency gains and performance outcomes. This contrast suggests that GenAI-supported writing and text-based production map more directly onto HSS assessment practices. At the same time, this close alignment intensifies concerns about intellectual agency and ethical use in HSS contexts.

HSS students also expressed significantly stronger concerns about GenAI-related risks. A notably larger proportion reported problems with accuracy (81.42%) and over-reliance (65.36%) compared to engineering students. These concerns reflect the interpretive and context-sensitive nature of HSS knowledge, where errors are often subtle and more consequential for argumentation and meaning-making. Ethical considerations were also much more prominent among HSS students (87.97%). In contrast, engineering students tended to prioritize technical accuracy, with ethical issues less directly embedded in routine problem-solving processes.

### *6.4. Sample imbalances*

Despite the relatively large sample size, a key limitation of this study lies in the substantial imbalance in sample composition. Female students account for 80.44% of the respondents. Although the gender distribution is broadly consistent with enrolment patterns in HSS disciplines in Chinese higher education, such disproportionality nevertheless may limit the generalizability of the findings beyond similarly structured populations.

More importantly, these imbalances may have influenced the magnitude and interpretation. In the case of gender differences, the overrepresentation of female students means that the overall estimates of perceived gains are more strongly shaped by female respondents' perceptions. As the analyses in Sections 5.2.8 show relatively small but consistent gender differences, particularly in cognitive and motivational outcomes, these effects may partly reflect differences in response tendencies, confidence framing, or technology-related self-efficacy rather than stable gender-based learning effects. Consequently, the observed gender gaps should be interpreted as indicative patterns within this specific sample structure rather than definitive evidence of population-level differences.

Similarly, students from education-related majors constitute 48.31% of the sample. This high proportion may have shaped the disciplinary comparisons. Given that nearly half of respondents belong to a single disciplinary cluster, the overall trends are likely weighted toward the experiences and pedagogical contexts typical of education majors. Although cross-disciplinary analyses in Section 5.2.7 suggest broadly consistent positive perceptions of GenAI across fields, the dominance of one discipline may have attenuated or amplified differences between groups, limiting the precision of inter-disciplinary contrasts.

While the results in this study provide meaningful insights into perceived learning impacts of GenAI in HSS contexts, the sample structure may introduce constraints on external validity. Future research should employ more balanced or stratified sampling strategies to enable stronger generalizations and more robust subgroup comparisons across gender and disciplinary categories.

### *6.5. Limitations and future research*

Despite its breadth, this study has several limitations. First, the use of self-reported data may introduce potential biases, as students can unintentionally overestimate or underestimate the learning benefits associated with AI use (Noroozi et al., 2025). Reflecting this concern, although students in our study reported substantial perceived improvements in academic performance, our findings indicated more ambiguous effects on independent thinking and creativity. This discrepancy suggests that perceived performance gains do not necessarily translate into meaningful cognitive development or deeper learning, which may ultimately constrain long-term learning outcomes. Future studies should triangulate self-reports with objective measures, such as platform usage logs, assessment outcomes, and instructor evaluations, to strengthen validity, potentially through embedded case studies.

Second, in this study, academic development is mainly understood as students' own perception of their achievement and success. Recent advances in predictive methods, particularly machine learning, make it possible to estimate student outcomes by analyzing behavioral patterns, engagement levels, and prior academic data. Such approaches may be considered in future studies to evaluate the impact of GenAI on students' academic development. However, the use of these predictive techniques comes with notable challenges and concerns. One key issue is algorithmic fairness, since models built on historical datasets may unintentionally reproduce or amplify existing disparities. This creates ethical questions around bias reduction and highlights the importance of thorough testing to promote fairness (Almalawi et al., 2024; Kesgin et al., 2025). Furthermore, many machine learning systems lack clarity in how their predictions are generated, making them difficult for both educators and students to interpret. This lack of transparency can weaken confidence in the results and raise accountability issues when such predictions influence educational decisions (Trejo-Macotela et al., 2026). Therefore, ensuring greater explainability and preserving human judgement remain crucial for the responsible and ethical use of predictive analytics in education (Borenstein et al., 2021; Kesgin et al., 2025).

Third, although differences across groups with varying durations of GenAI use were analyzed to explore duration-related differences in perceived outcomes, these groups comprised different participants, which limited causal interpretation. Such comparisons do not capture within-individual change over time and therefore cannot establish longitudinal developmental trajectories. In future studies, longitudinal designs tracking the same cohort over time are needed to more rigorously capture trajectories of GenAI adoption, skill development, and career preparedness.

Lastly, the exclusive use of an anonymized questionnaire constrains interpretive depth. Future work can adopt mixed-method approaches, particularly qualitative interviews, to provide deeper insight into how students negotiate tensions between efficiency, independent thinking, creativity, and academic responsibility when engaging with GenAI. While this will require more effort, it will improve the validity of the research findings, thus contributing to a more comprehensive and accurate understanding of how GenAI impacts HSS education.

## 7. Conclusion

This study investigated the use of GenAI among Chinese HSS students, examining its learning impacts, challenges, and disciplinary implications through a large-scale survey. The findings show that GenAI has been widely adopted for writing support, literature searches, conceptual clarification, and exploratory idea generation. High reported gains in learning efficiency (by 96.07%) and active learning motivation (by 65.03%) suggest that GenAI effectively supports knowledge construction and promotes student engagement.

GenAI also influences higher-order cognition. Over half of respondents reported gains in independent thinking and creativity, suggesting its capacity to support advanced intellectual work. However, many others reported little improvement or even declines, highlighting the need for reflective, self-regulated use to ensure GenAI strengthens rather than displaces intellectual agency. In terms of gender, male students reported greater perceived gains in these areas than female students. In addition, differences observed across groups with varying durations of GenAI use suggest a potential pattern in which deeper educational benefits are more strongly reported among more experienced users; however, this should be interpreted as an association rather than a confirmed developmental trajectory. Disciplinary differences were also evident. Students in arts and in economics and management reported stronger gains in independent thinking, creativity, and learning motivation than those in education and law. These patterns suggest the importance of sustained engagement and tailored GenAI literacy training to position GenAI as a long-term, domain-adapted cognitive partner rather than a short-term, uniform productivity tool.

Most students also perceived improvements in their academic performance, a perception that aligns with their evaluation of GenAI as highly aligning to their HSS disciplines. However, these positive evaluations coexisted with frequent experiences of inaccurate outputs and strong demands for improved accuracy. This tension suggests that perceived disciplinary adaptability may be superficial and that reported performance gains may, in part, reflect limitations in current assessment practices to capture students' genuine learning and analytical depth.

These findings on learning and performance should be interpreted cautiously because they rely on self-reported data, which may be affected by recall bias and students' tendency to overestimate or underestimate the educational benefits of GenAI use. The discrepancy between perceived academic improvement and the less consistent effects on independent thinking and creativity further suggests that subjective perceptions may not accurately reflect deeper cognitive development. Future research should therefore incorporate objective measures, such as assessment outcomes, usage analytics, and instructor evaluations, to provide a more comprehensive understanding of GenAI's impact on learning and performance.

A large majority of respondents (87.97%) rated ethical considerations as important or very important, highlighting the central role ethics plays in students' engagement with GenAI. This finding aligns with the HSS disciplines' strong emphasis on authorship, argumentation, and intellectual responsibility. Although more than half of respondents reported satisfaction with the privacy performance of GenAI tools, a considerable proportion (37.27%) remained neutral, indicating ongoing uncertainty and highlighting the need for GenAI systems that are more transparent and ethically aligned.

With respect to institutional curriculum integration, students clearly favored partial or optional incorporation of GenAI into the curriculum, accompanied by practice-oriented training, explicit guidelines, and discipline-specific policies. They prioritized enhancements in output accuracy, analytical depth, literature search capabilities, and integration with professional software, signaling a strong demand for academically robust and versatile GenAI systems rather than generic conversational tools.

Our findings indicate that GenAI holds significant potential to enhance learning in HSS education. However, these benefits are not automatic. They are maximized when GenAI use is embedded within ethical frameworks, reflective practice, self-regulated learning strategies, and discipline-sensitive pedagogical scaffolding. To realize this potential, higher education institutions should move beyond ad-hoc adoption and implement targeted training, clear governance structures, and thoughtfully designed integration strategies that support both academic integrity and the development of responsible, innovative, and professionally prepared graduates.

**Data availability:** Data will be made available on reasonable request to the corresponding author.

**Competing interest:** The authors declare no competing interests.

**Ethics statement:** The study protocol for the user survey was reviewed and approved by the University Research Ethics Review Panel. All procedures were conducted in accordance with applicable ethical guidelines and regulations.

**Informed consent**: Informed consent was obtained from all individual participants for participation in this study. Anonymity of their identification and voluntary participation were assured. No identifiable data of participants were present within the manuscript.

**Funding:** The research received no funding.